\documentclass[conference]{IEEEtran}
\newcommand{\bigrho}{\makebox{\Large\ensuremath{\rho}}}

\usepackage{color}

\usepackage{cite}
\usepackage[pdftex]{graphicx}
\usepackage[ruled,vlined]{algorithm2e}
\usepackage{algpseudocode}
\usepackage{mathtools, cuted}
\usepackage{amsmath}
\usepackage{amsthm}
\usepackage{bm}
\theoremstyle{definition}

\usepackage{multicol}
\usepackage{amsfonts}
\usepackage{xcolor}
\usepackage{subcaption}
\usepackage{caption}
\usepackage{caption}
\usepackage{subcaption}
\usepackage{multirow}
\usepackage{mathrsfs}

\usepackage[margin=0.8in, top=0.92in,bottom=1.1in, columnsep=0.24in]{geometry}

\ifCLASSINFOpdf
\else
\fi
\begin{document}
%
\title{Study on Downlink CSI compression: Are Neural Networks the Only Solution?}



\author{\IEEEauthorblockN{K. Sai Praneeth\IEEEauthorrefmark{1}, Anil Kumar Yerrapragada\IEEEauthorrefmark{2}, 
Achyuth Sagireddi\IEEEauthorrefmark{3},
Sai Prasad\IEEEauthorrefmark{4}
and Radha Krishna Ganti\IEEEauthorrefmark{5}}
\IEEEauthorblockA{Department of Electrical Engineering\\
Indian Institute of Technology
Madras \\ Chennai, India  600036\\
Email: \IEEEauthorrefmark{1}praneethk@smail.iitm.ac.in,
        \{\IEEEauthorrefmark{2}anilkumar,
        \IEEEauthorrefmark{3}achyuth,
        \IEEEauthorrefmark{4}venkatasiva\}@5gtbiitm.in,
		\IEEEauthorrefmark{5}rganti@ee.iitm.ac.in
}
}


%


\maketitle

\begin{abstract}

Massive Multi Input Multi Output (MIMO) systems enable higher data rates in the downlink (DL) with spatial multiplexing achieved by forming narrow beams. 
The higher DL data rates are achieved by effective implementation of spatial multiplexing and beamforming which is subject to availability of DL channel state information (CSI) at the base station. For Frequency Division Duplexing (FDD) systems, the DL CSI has to be transmitted by User Equipment (UE) to the gNB and it constitutes a significant overhead which scales with the number of transmitter antennas and the granularity of the CSI.
To address the overhead issue, AI/ML methods using auto-encoders have been investigated, where an encoder neural network model at the UE compresses the CSI and a decoder neural network model at the gNB reconstructs it. However, the use of AI/ML methods has a number of challenges related to (1) model complexity, (2) model generalization across channel scenarios and (3) inter-vendor compatibility of the two sides of the model. In this work, we investigate a more traditional dimensionality reduction method that uses Principal Component Analysis (PCA) and therefore does not suffer from the above challenges. Simulation results show that PCA based CSI compression actually achieves comparable reconstruction performance to commonly used deep neural networks based models.

\end{abstract}


\IEEEpeerreviewmaketitle

\section{Introduction}
In OFDM-based MIMO systems, with Time Division Duplexing (TDD) mode of operation, channel reciprocity is used to deduce Downlink Channel State Information (CSI) from the Uplink channel characteristics. 
However for Frequency Division Duplexing (FDD) based systems, the downlink CSI has to be transmitted by the UE to the base station (sometimes referred to as network).
This feedback typically requires huge uplink resources in order to transfer the whole CSI. To reduce overhead on the uplink, 3GPP has introduced codebook-based methods which allow partial CSI feedback to be conveyed, helping the base station to understand the channel conditions. 
In \cite{38214_3gpp}, the Type-I codebook was developed, which communicates only wide-band channel information. This was followed by the Type-II codebook, which communicates higher resolution of CSI feedback such as sub-band level channel information along with wide-band information.
Further improvements were added in \cite{38214_3gpp_r16} with the introduction of the enhanced Type II (eType-II) codebook, which reduces overhead by compressing the sub-band level channel information using a DFT based transformation. The Type-I codebook has the least overhead but suffers from performance degradation \cite{10339312} compared to Type-II and eType-II. The Type-II and eType-II codebooks allow higher granularity of CSI reporting but at the cost of higher overheads. \cite{38214_3gpp_r18} supports Doppler codebooks which are applicable to high mobility scenarios but also suffer from higher overhead. 

\begin{figure}[ht!]
    \centering
    \includegraphics[width = 0.48\textwidth]{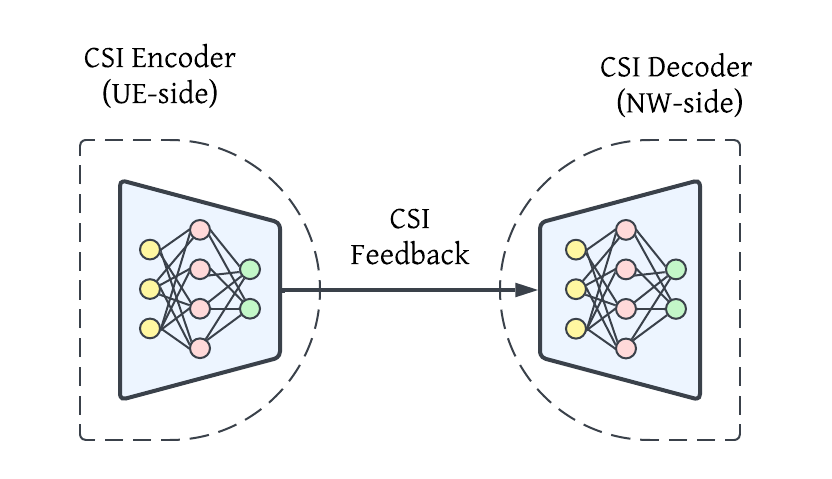}
    \caption{AI/ML induced CSI compression framework over-view.}
    \label{fig:architecture_csi_overview}
\end{figure}

To address the increased overhead issues, \cite{38214_3gpp_r18} introduced the study of AI/ML based CSI compression, which essentially uses neural networks to compress the channel at the UE and re-construct it at the base station as depicted in Figure \ref{fig:architecture_csi_overview}. Deep learning techniques such as auto-encoders, have obtained considerable attention because of their potential to further reduce the overhead and optimize CSI compression~\cite{8918798,wen2018deeplearningmassivemimo,9768327,9296555,9466243,9171358,9473840,9538824,9126231,9446900,9417115,9481880,8744528}.  

Despite the promise shown by AI/ML for CSI compression, there are several issues related to their practical deployment. In this paper: 
\begin{itemize}
    \item  We investigate the necessity of neural network-based CSI compression by considering factors like computational complexity, generalization across channel scenarios, and inter-vendor compatibility of the two sides of the CSI compression models. 
    \item We propose a PCA based method which does not suffer from the above issues. Using two representations of the channel (angular-delay domain and eigenvector) we compare the performance of PCA based compression with state-of-art neural networks.
\end{itemize}


\section{System Model and PCA based compression}\label{sectionII}
In this work, we consider two different representations of the wireless channel i.e., $1)$ Angular-Delay domain representation and $2)$ Eigenvector representation. In this section, we explain the channel modeling followed by the Principal Component Analysis (PCA) for both domains. 

\subsection{System Model}
We consider an $N$ sub-carrier OFDM system with $N_t$ transmit antenna ports and $N_{rx}$ receive antennas. Without loss of generality, we assume that the base station is the transmitter and the UE is the receiver. 
The received signal at the $s^{th}$ sub-carrier $\mathbf{Y}[s]$ is given by

\begin{equation} \label{eq:1}
    \mathbf{Y}[s] = \mathbf{H}[s]\mathbf{x}[s]  + \mathbf{W}[s],
\end{equation}

where $\mathbf{Y}[s]$ is of size $N_{rx}\times 1$. $\mathbf{H}[s]$ is the $N_{rx}\times N_{t}$ channel, $\mathbf{x}[s]$ is the $N_{t}\times 1$ transmitted sequence and $\mathbf{W}[s]$ is the $N_{rx}\times 1$ noise vector.

In this paper we consider perfect knowledge of channel. We also assume that $N_{rx} = 1$. In this case, the channel across the entire bandwidth (all $N$ sub-carriers) can be represented by the $N\times N_{t}$ matrix $\mathbf{H_f}$ given by,
\begin{equation}\label{freq_spatial}
    \mathbf{H_f} = 
    \begin{bmatrix}
  \mathbf{h}[1] \\
  \mathbf{h}[2] \\
  \vdots \\
  \mathbf{h}[N] \\
\end{bmatrix}
\end{equation}
where $\mathbf{h}[\cdot]$ is the $1\times N_{t}$ representation of $\mathbf{H}[\cdot]$.
\subsection{Angular-Delay (AD) Domain Data}
The spatial-frequency domain channel matrix $\mathbf{H_f}$ can be sparsed in the angular-delay domain using a 2D discrete Fourier transform (DFT) as given in~\cite{wen2018deeplearningmassivemimo}.

\begin{equation}\label{angle_delay_eq}
    \mathbf{H_{ad}} = \mathbf{F_d} \mathbf{H_f} \mathbf{F_a}^H,
\end{equation}
where $\mathbf{F_d}$ and $\mathbf{F_a}$ are $N \times N$ and $N_t \times N_t$ DFT matrices respectively.

It is important to note that only a few rows, say $L$, in $\mathbf{H_{ad}}$ are significant and all other rows would have values close to zero. Thus, by selecting the significant rows, the final angular-delay representation of channel is given by 
\begin{equation}\label{final_angle_delay_eq}
    \mathbf{H_{t_L}} \in \mathbb{C}^{L \times N_t} 
\end{equation}
$\mathbf{H_{t_L}}$ can be interpreted as a time domain channel with $L$ taps. 

The Principal Component Analysis of the angular-delay domain channel involves finding the independent time-domain channels across all the transmitter antennas i.e, the angular domain. 
The PCA on the antenna dimension determines the minimum number of components (antennas with unique channel properties are chosen whereas those antennas with redundant channel information are ignored) that capture the angular properties of the channel. 

This technique, when applied at the UE, performs compression and is analogous to an encoder neural network. The task at the base-station would be to perform the inverse PCA to recover the angular-delay representation of the channel. The compression ratio scales with the choice of number of delays/taps and the number of principal components.
Note that the transformation of PCA would be different for each instance of the channel and each UE. 
The initial reduction of the channel $\mathbf{H_{f}}$ to $L\times N_t$, originally from $N\times N_t$, is achieved by the transformations in Eq.~\eqref{angle_delay_eq} and further reduction in the spatial dimension from $N_t$ to $N_t'$, is achieved by the implementation of PCA. 

The overall CSI feedback with PCA is $\mathbf{H}_{PCA,AD}$ of dimension $L\times N_t'$ along with a transformation matrix $\mathbf{H}_{N_t'}$ of dimension $N_t'\times N_t$ to aid in inverse PCA at the network side. 
The overhead reduction of PCA for angular-delay domain data $OR_{PCA,AD}$ is given as
\begin{equation} \label{eq:3}
    {OR_{PCA,AD} = \frac{(LN_t) - (L+N_t)N_t'}{LN_t}}
\end{equation}

It can be observed from Eq.~\eqref{eq:3} that the compression ratio is scaled with the number of principal components $N_t'$ chosen for a given $L$ and $N_t$. 

At the network side, the inverse PCA, performed using the two received matrices is given by
\begin{equation} \label{eq:4}
    \mathbf{\hat{H}}_{PCA,AD}  = \mathbf{\Tilde{H}}_{PCA,AD} \big(\mathbf{\Tilde{H}}_{N_t'}\big)^H + Q_L,
\end{equation}
where $\mathbf{\Tilde{H}}_{PCA,AD}$ and $\mathbf{\Tilde{H}}_{N_t'}$ are the received PCA-based compressed channel and received PCA transformation matrices respectively and $Q_L$ is the loss incurred due to quantization.

$\mathbf{\hat{H}_{PCA,AD}} $ is the reconstructed angular-delay channel at base station.
The reconstructed spatial-frequency domain channel can be obtained by performing inverse operations to the transformations given in Eq.~\eqref{angle_delay_eq} as 

\begin{equation} \label{eq:5}
    \mathbf{\hat{H}_{f}} = \mathbf{F_d}^H \mathbf{\hat{H}_{PCA,AD} } \mathbf{F_a}
\end{equation}

We use an approximated version of Generalized Cosine Similarity (GCS) as a metric to evaluate the closeness of reconstructed channel matrix $\mathbf{\hat{H}_{f}}$ and the true channel $\mathbf{H_{f}}$. The approximated GCS is given by,
\begin{equation} \label{eq:6}
    {\text{GCS} = \bigrho_{AD} = \frac{ | \mathbf{\hat{H}_f}^{H} \mathbf{H_f}|}{\lVert\mathbf{\hat{H}_f}\rVert_2 \lVert \mathbf{{H}_f}\rVert_2}}.
\end{equation}

\subsection{Eigenvector (EV) Data}
For the eigenvector representation of the channel, we divide the total bandwidth ($N$ sub-carriers) into $N_{SB}$ sub-bands. In this paper we assume that a sub-band consists of $4$ resource blocks. The sub-band channel matrix $\mathbf{H}_k$ is the average of channel matrices of all Resource Elements (REs) of $k_{th}$ sub-band. The eigenvector data is generated by computing the Eigen Value Decomposition (EVD) of $ \mathbf{H}_k^H \mathbf{H}_k$. Concatenating the eigenvectors of each sub-band as given below we obtain
\begin{equation} \label{eq:7}
     \mathbf{H}_{EV} = [\mathbf{E}_1,\mathbf{E}_2,\hdots,\mathbf{E}_{N_{SB}} ],
\end{equation}
where, $\mathbf{E}_k$ ($k=1,2\hdots N_{SB}$), is an $N_t\times R$ matrix. 
Note that $N_t$ is number of transmitter antenna ports, $R$ is the rank of sub-band channel $\mathbf{H}_k$. Assuming $N_{rx} = 1$, the rank $R$ is $1$. Therefore the dimension of $\mathbf{H}_{EV}$ is $N_t \times N_{SB}$.
 
For $\mathbf{H}_{EV}$ as given by Eq.~\eqref{eq:7}, we perform PCA on the sub-band dimension $N_{SB}$. 
The idea is to select unique eigenvectors across all the sub-bands.
Similar to the angular-delay domain data, the reconstruction of eigenvector data at the base station is performed by making use of the $N_{SB}\times N_{SB}'$ PCA transformation matrix $\mathbf{\Tilde{H}}_{N_{SB}'}$, transmitted along with the PCA compressed eigenvector data $\mathbf{\Tilde{H}}_{PCA,EV}$ as feedback from the user. Here $N_{SB}'$ is the number of significant principal components.
The reconstructed eigenvector data $\hat{\mathbf{H}}_{EV}$ at the base station is given by 
\begin{equation} \label{eq:9}
    {\mathbf{\hat{H}}}_{EV} = \mathbf{\tilde{H}}_{PCA,EV} \big(\mathbf{\tilde{H}}_{N_{SB}'}\big)^H + Q_L,
\end{equation}
where $\mathbf{\tilde{H}}_{PCA,EV}$ and $\mathbf{\tilde{H}}_{N_{SB}'}^{bs}$ are received PCA-based compressed channel and received PCA transformation matrices respectively and $Q_L$ is the quantization loss.

Thus the overhead reduction for the eigenvector data $OR_{PCA,EV}$ is given as 
\begin{equation} \label{eq:CR_EV_PCA}
    {OR_{PCA,EV} = \frac{(N_{SB} N_t) - (N_{SB}+N_t)N_{SB}'}{N_{SB}N_t}}.
\end{equation}
where, $N_{SB}'$ is the number of principle components in the sub-band dimension. The closeness of the true channel and the reconstructed channel is computed using an approximated cosine similarity as given by,
\begin{equation} \label{eq:6}
    {\text{GCS} = \bigrho_{EV} = \frac{ |\mathbf{\hat{H}}_{EV}^H \mathbf{H}_{EV} |}{\lVert\mathbf{\hat{H}}_{EV}\rVert_2 \lVert \mathbf{ {H}}_{EV}\rVert_2}}.
\end{equation}

For both angular-delay domain and eigenvector representation of the channel, the PCA based CSI compression requires us to choose the right number of principal components $N_t'$ and $N_{SB}'$ respectively. 
The following section, describes the prominent neural networks used for CSI compression with which we make a comparison with our PCA based compression method.

\subsection{Feedback bits required}
This sub-section explores the total number of feedback bits needed to perform PCA based CSI compression. We define the total feedback bits $B_T$ needed, for sending PCA based CSI feedback for both angular-delay and eigenvector data as follows,

\begin{align*}\label{eq:13}
    B_T &= B_C + B_R \left(\frac{\tau_{p}}{\tau_r} \right),\\
    &= \left( LN_t' + N_t'N_t \left(\frac{\tau_{p}}{\tau_r}\right)\right)\left(2Q\right) \text{ - for AD data}\\
    &= \left( N_t N_{SB}' +N_{SB}' N_{SB} \left( \frac{\tau_{p}}{\tau_r} \right)\right)\left(2Q\right)\text{ - for EV data}
\end{align*}
where, $B_C$ represents the compressed bits, $B_R$ represents the bits needed to perform reconstruction of the channel at the network side, $\tau_p$ represents the CSI reporting periodicity, $\tau_r = k\tau_p$, for $k=1,2,3\hdots,$ represents the periodicity at which reconstruction bits $B_R$ are fed-back to the network from UE, $Q$ represents the quantization bits and factor $2$ indicates the real and imaginary split of the complex numbers.

\section{AI/ML for CSI compression}\label{sec: ai_ml}
 We study two architectures: CSINet~\cite{wen2018deeplearningmassivemimo} and EVCSINet~\cite{9538824} which work on angular-delay domain data and eigenvector data respectively. 

\subsection{CSINet}
For the CSINet model we use the same architecture described in~\cite{wen2018deeplearningmassivemimo}. The encoder has a series of convolutional layers for feature extraction followed by a dense layer for feature compression. The decoder has a dense layer to decompress the features followed by a ResNet \cite{resnet_arch} like architecture to generate the reconstructed channel from the features.  

\subsection{EVCSINet}
For the EVCSINet model we use the same architecture described in~\cite{9538824}. The encoder is based on fully connected layers to learn a lower dimensional representation of the eigenvectors. The decoder is a ResNet \cite{resnet_arch} like architecture for the reconstruction of the eigenvectors. We note that while the decoder architecture in~\cite{9538824} contains 27 convolutional blocks, in this paper we use a lighter version with only 15 convolutional blocks. 

\subsection{Data Generation}
For simulation purposes we use two types of data. The first is from publicly available datasets of channel scenarios like CDLA-30 and CDLA-300~\cite{Oppo}. Additionally we also use our own channel data of the Urban Macro (UMa) scenario. Our datasets are generated using QuaDRiGa~\cite{Quadriga}, a MATLAB based software tool for developing 3GPP compliant channels. 
\renewcommand{\arraystretch}{1.1}
\begin{table}
\caption{Simulation parameters for generation of a private data using Quadriga.}
\centering
\begin{tabular}{ |l||c|  }
 \hline
\textbf{Parameter}  & \textbf{Value} \\\hline
  Scenario  & Urban Macro (UMa) \\
  Center Frequency & $2$ GHz \\
  Bandwidth & $10$ MHz \\
    Sub-carrier spacing (SCS) & $15$ KHz \\
    Number of Physical Resource Blocks & $52$ \\
    Number of Resource Elements ($N$) & $624$ \\
    Number of sub-bands ($N_{SB}$) & $13$ \\
    Number of Transmit Antennas at BS ($N_{tx}$) & $16$ \\
    Number of CSI-RS ports ($N_t$) & $32$ \\
    Number of Receiver Elements at UE ($N_{rx}$) & $1$ \\
    Antenna panel dimensions of BS & $2 \times 8 \times 2 \times 1 \times 1$ \\
    Antenna panel dimensions of UE & $1 \times 1 \times 1 \times 1 \times 1$         \\
    Cell radius & $100$ m \\
    Number of sectors  & $3$ \\ \hline
\end{tabular}
\label{tab:simulation parameters}
\end{table}

The data generation for our UMa data is based on the simulation parameters defined in Table~\ref{tab:simulation parameters}. Using Quadriga, we place users uniformly across a $100$m, 3-sector site. Quadriga generates the Channel Impulse Response (CIR) between each UE and the base station located at the center of the site. By applying a DFT to the CIR, the Channel Frequency Response (CFR) is derived, capturing the effects of multipath delays in the channel. This CFR data is then used to generate two distinct types of wireless channel data.

\subsubsection{Angular-Delay domain data}
To generate the angular-delay data, we multiply the CFR (Eq.~\eqref{freq_spatial}) with DFT matrices as indicated in Eq.~\eqref{angle_delay_eq}. We then consider only the first $L$ significant rows to obtain $\mathbf{H_{t_L}}$ indicating that the channel has $L$ significant paths. The values of $L$ are given in Table~\ref{table:L_NSB_Channels}. The CSINet model input is derived from $\mathbf{H_{t_L}}$ with the real and imaginary parts split such that model input is of size $L\times N_t\times 2$.  

\subsubsection{Eigenvector data}
The first step in generating eigenvector data is performing sub-band level averaging of the CFR in each of the $N_{SB}$ sub-bands. This involves grouping the channels of each resource element within each sub-band and averaging them. The average channel in each sub-band is then decomposed using EVD, as shown in Eq.~\eqref{eq:7}. Then the selection of top eigenvectors is based on the rank of the channel matrix, which is determined by the minimum of $N_t$ and $N_{rx}$. In our case, since $N_{rx}=1$, we choose the top eigenvector in each sub-band. For input to EVCSINet, the eigenvectors across all sub-bands are concatenated followed by a split of the real and imaginary parts. The split is such that all the real values of eigenvectors of all sub-bands appear together followed by all the imaginary parts.

\renewcommand{\arraystretch}{1.3}
\begin{table}
\caption{Choice of $L$ and $N_{SB}$ for various channels for Angular-Delay and Eigen Vector data respectively}
\centering
\begin{tabular}{ |c||c||c| }
 \hline
\textbf{Channel} & \textbf{Angular-Delay data} & \textbf{Eigen Vector data} \\
 & \textbf{\# Taps ($L$)} & \textbf{\# Sub-Bands ($N_{SB}$)}\\ \hline
CDLA-30 & 5 & 12 \\
CDLA-300 & 25 & 12\\
Own Data & 25 & 13\\\hline

\end{tabular}\label{table:L_NSB_Channels}
\end{table}


\section{Performance comparison and Results} \label{sec:V}
We compare the compression accuracy of PCA based CSI feedback with the well-known CSINet and EVCSINet architectures. 
In this work, we consider three different datasets to evaluate the effectiveness of CSI compression with PCA-based method and AI/ML methods mentioned above. We use two publicly available datasets along with one of our own generated dataset as discussed in Section \ref{sec: ai_ml}. 

\subsection{CSI compression : PCA vs AI/ML}
\begin{figure}[h!]
    \captionsetup{justification=centering}
     \centering
     \begin{subfigure}[b]{0.5\textwidth}
         \centering
         \includegraphics[width=\textwidth]{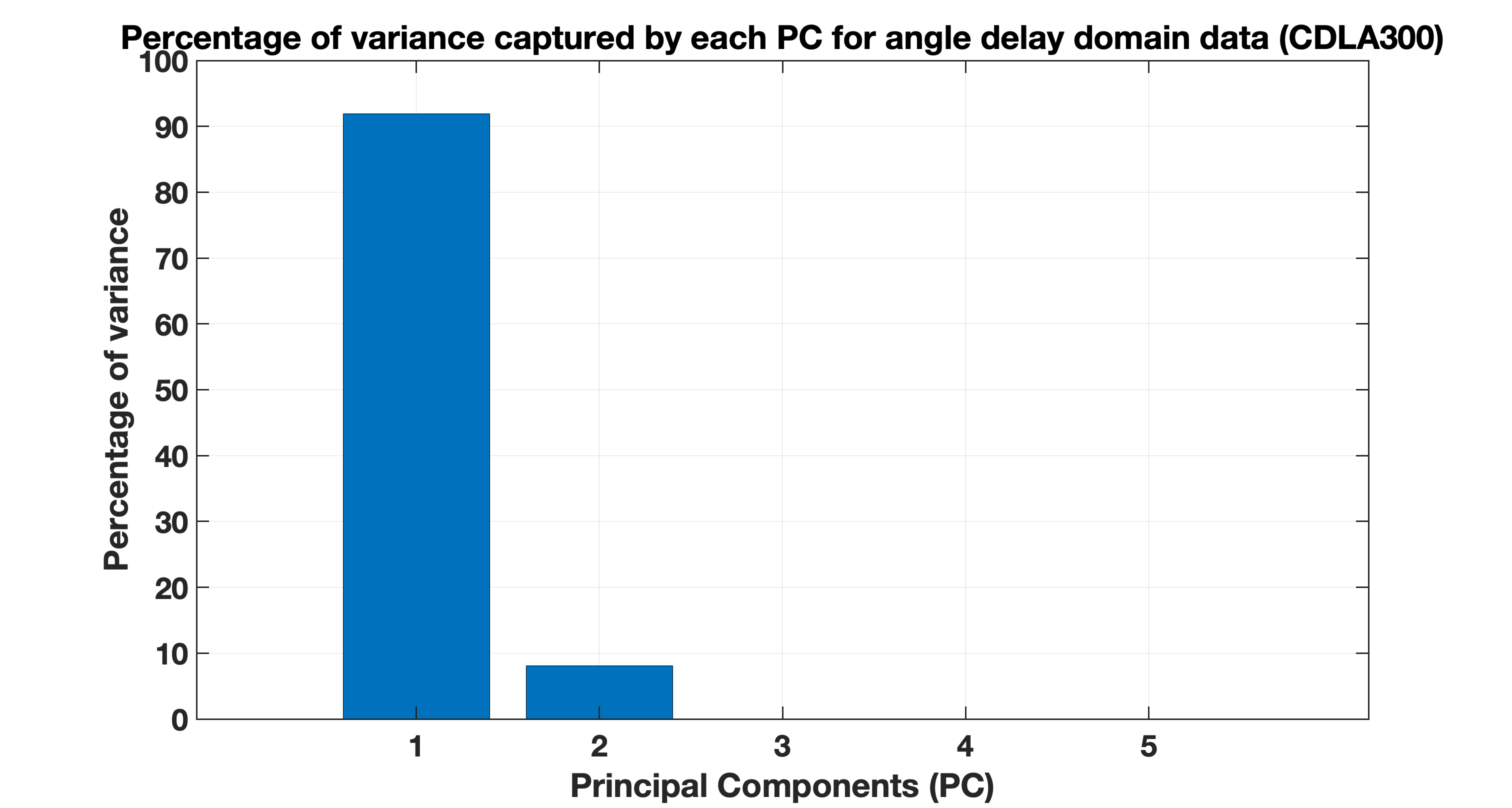}
         \caption{}
         \label{fig:accuracy_vs_epoch}
     \end{subfigure}
     \\
     \begin{subfigure}[b]{0.5\textwidth}
         \centering
         \includegraphics[width=\textwidth]{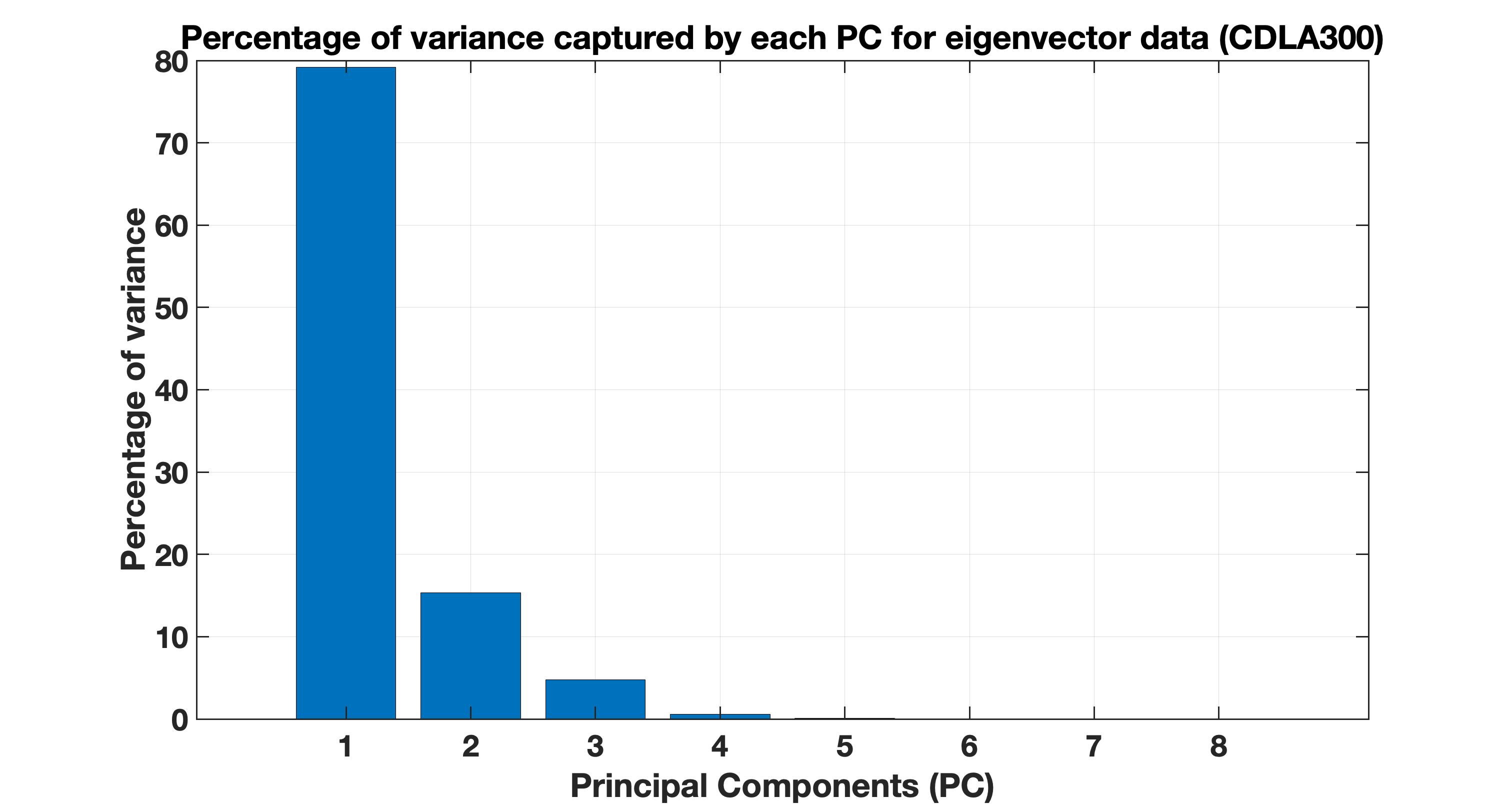}
         \caption{}
         \label{fig:loss_vs_epoch}
     \end{subfigure}
        \caption{Percentage of variance captured by each principal component for CDLA300 channel represented as (a) Angle Delay Domain data and (b) Eignevector data}
        \label{fig:accuracy_loss_vs_epoch}
\end{figure}

For the angular-delay domain data of dimension $L\times N_t$, we perform PCA on the transmit antenna ports (CSI-RS ports) dimension $N_t$ to determine the principal components with significant variance. Similarly for eigenvector data we perform PCA on the sub-band dimension of the channel.
For the different data sets considered, i,e., CDLA-30, CDLA-300, and (our) UMa, it is interesting to note that $99\%$ of the channel instances require only $2$ principal components for angular-delay data and $3$ principal components for eigenvector data to capture most of the variance (as illustrated in Figure~\ref{fig:accuracy_loss_vs_epoch}).

\renewcommand{\arraystretch}{1.7}
\begin{table*}
\caption{Results comparing AI/ML CSINet with PCA}
\centering
\begin{tabular}{|*{20}{c|}}
\hline

\multicolumn{2}{|c|}{\textbf{Model/Dataset}}  & \multicolumn{1}{|c}{\textbf{Train : CDLA-30}} & \multicolumn{1}{|c}{\textbf{Train : CDLA-300}} & \multicolumn{1}{|c|}{\textbf{Train : Own data}} \\ \hline

 \textbf{Model} & \textbf{Parameter} & \textbf{Test : CDLA-30} & \textbf{Test : CDLA-300} & \textbf{Test : Own data}  \\ \hline
CSINet  & Cosine Similarity & 0.9973 & 0.9707 &  0.9322 \\ 

& Overhead Reduction (\%) - $OR_{CSINet}$&  77&  93&  93\\
\hline
PCA with $N_t'=1 $& Cosine Similarity & 0.8943 & 0.9384 &  0.8492  \\ 

& Overhead Reduction (\%) - $OR_{PCA,AD}$ &  77&  93&  93\\ 
\hline





PCA with $N_t'=2$ & Cosine Similarity & 0.8944 & 0.9796 & 0.9404  \\ 

& Overhead Reduction (\%) - $OR_{PCA,AD}$ &  54&  86&  86\\ 
\hline




PCA with $N_t'=3$& Cosine Similarity & 0.8944 & 0.9798 & 0.9687 \\ 

& Overhead Reduction (\%) - $OR_{PCA,AD}$ &  31&  79&  79\\
\hline




\end{tabular}\label{table:CSINet_vs_PCA}
\end{table*}

\renewcommand{\arraystretch}{1.7}
\begin{table*}
\caption{Results comparing AI/ML EVCSINet with PCA}
\centering
\begin{tabular}{|*{20}{c|}}
\hline

\multicolumn{2}{|c|}{\textbf{Model/Dataset}}  & \multicolumn{1}{|c}{\textbf{Train : CDLA-30}} & \multicolumn{1}{|c}{\textbf{Train : CDLA-300}} & \multicolumn{1}{|c|}{\textbf{Train : Own data}} \\ \hline

 \textbf{Model} & \textbf{Parameter} & \textbf{Test : CDLA-30} & \textbf{Test : CDLA-300} & \textbf{Test : Own data}  \\ \hline

EVCSINet  & Cosine Similarity & 0.9865 & 0.9859 &  0.9172 \\ 

& Overhead Reduction (\%) - $OR_{EVCSINet}$ &  89&  89&  89\\ 
\hline
PCA with $N_{SB}'=1 $& Cosine Similarity & 0.9539 & 0.9526 & 0.8508   \\ 

&  Overhead Reduction (\%) - $OR_{PCA,EV}$ &  89&  89&  89\\ 
\hline


PCA with $N_{SB}'=2$ & Cosine Similarity & 0.9727 & 0.9716 &  0.9327 \\ 

& Overhead Reduction (\%) - $OR_{PCA,EV}$ &  77&  77&  78\\ 
\hline




PCA with $N_{SB}'=3$& Cosine Similarity & 0.9730 & 0.9719 & 0.9539 \\ 

& Overhead Reduction (\%) - $OR_{PCA,EV}$ &  65&  65&  67\\ 
\hline




\end{tabular}\label{table:EVCSINet_vs_PCA}
\end{table*}


Our evaluations show that for the angular-delay domain data, the choice of number of channel taps is crucial for PCA to achieve comparable compression performance compared to CSINet. With sufficient number of channel taps (as indicated in Table \ref{table:L_NSB_Channels}), PCA is able to achieve similar CSI compression performance to that of CSINet, as given in the last two columns of Table \ref{table:CSINet_vs_PCA}.
With eigenvector data, for CDLA-30 and CDLA-300 datasets, PCA with $N_{SB}'=1$ principal components is able to achieve almost similar CSI compression metrics to that of EVCSINet with same overhead reduction as shown in  Table \ref{table:EVCSINet_vs_PCA}. In the case of UMa, PCA with $N_{SB}'=2$ offers comparable results with EVCSINet with a slight increase in overhead.


    \subsection{Generalisation and Vendor Inter-operability} It is a known fact that one of the major drawbacks of AI/ML models is the lack of generalization. In some cases where training is done using data from a specific channel scenario and inference is performed on data from another channel scenario, the model finds it difficult to generalize and the performance drops significantly. For example, our experiments show that with angle-delay domain data, the CSINet model trained on CDLA300 channels and tested on UMa Channels shows a cosine similarity of only $55\%$. This is in contrast to the $99\%$ when model is tested on the same scenario as that of the training.  Such generalization issues lead to developing a large number of AI/ML models to cater different cells and scenarios. 
    
    Another drawback for AI/ML models is the issue of inter vendor compatibility.
    There are multiple UE and base station vendors, each of which could develop proprietary models for CSI compression. 
    With PCA-based CSI feedback, the compressed channel and the transformation matrix required for reconstruction are transmitted in every CSI report instance.
    This way, there is no fixed compression matrix at the UE and no fixed reconstruction matrix at the base station. This eliminates the need for vendor inter-operability.
    Since the compression matrix and reconstruction matrix are computed for each instance of CSI report, there is no need for generalization across multiple cells and scenarios. 
    Thus, with the PCA-based approach, we don't encounter the issues of generalization and inter-operability.

\section{Conclusion and Further Work}\label{sec:VI}
In this work, we compared the compression efficiency and overhead reduction of deep neural network based CSI feedback and a conventional machine learning approach, PCA. The PCA method is a linear dimensionality reduction method that helps us to find the minimum number of principal components required to represent the maximum variance of the data. We considered two different representations of the wireless channel i.e., $1)$ Angular-Delay domain representation $2)$ Eigenvector representation. 
The deep neural network model architectures considered are CSINet and EVCSINet and these models are trained with angular-delay domain channel data and eigenvector channel data respectively.

Based on our results, PCA based CSI compression achieves almost similar CSI compression metrics as compared to neural networks. Additionally PCA based CSI compression doesn't suffer from issues like generalisation and vendor inter-operability.
Thus, we suggest that PCA based CSI compression can also be considered as a choice for CSI compression. 
We would like to further investigate the PCA-based approach with all 3GPP channel models and scenarios to verify if the observations drawn here hold true. Additionally we would like to employ non-linear dimensionality reduction techniques such as manifold learning to compress CSI feedback and thus reduce CSI-related overhead on the uplink.

 \section*{Acknowledgment}
This work was funded by MEiTY, Government of India, through the project, “Next Generation Wireless Research and Standardization on 5G and Beyond”, by the Department of Telecommunications (DoT), Government of India, through the 5G testbed project, and by ANSYS Software Pvt. Ltd. through their Doctoral Fellowship award program. 



\bibliographystyle{IEEEtran}
\bibliography{bibfile}

\end{document}